\documentclass{jpsj3}
\usepackage{graphicx}
\usepackage{dcolumn}
\usepackage{bm}
\usepackage{color}

\title{%
    Effect of Coulomb Interaction and Disorder
 on Density of States
 in Conventional Superconductors
}
\author{%
Takanobu Jujo
\thanks{E-mail address: jujo@ms.aist-nara.ac.jp}
}
\inst{%
Division of Materials Science, 
Graduate School of Science and Technology, 
Nara Institute of Science and Technology, 
Ikoma, Nara 630-0101, Japan
}
\abst{%
  The density of states of the disordered s-wave superconductor is
  calculated perturbatively.
  The effect of Coulomb interaction on diffusively moving electrons
  in the normal state has been known before,
  but in the superconducting state both diffuson
  and the screened Coulomb interaction are modified.
  Therefore, the correction to the density of states
  in the superconducting state exhibits an energy dependence
  different from that of the normal state.
  There is a dip structure in the correction part
  because the interaction has a peak at twice the energy
  of the superconducting gap.
  The Coulomb interaction and the superconducting fluctuation
  cannot be treated separately because
  the density fluctuation is coupled to the phase fluctuation
  in the superconducting state. 
  This coupling results in 
  the absence of divergence around the gap edge in the correction part,
  which suggests the validity of this perturbation calculation.
}

\begin{document}
\setlength{\textwidth}{504pt}
\setlength{\columnsep}{14pt}
\hoffset-23.5pt
\maketitle

\section{Introduction}

The conventional s-wave superconductor is
not affected by the impurity scattering itself
because nonmagnetic impurities do not break
the symmetry of s-wave superconductors.
~\cite{anderson59}
In general there exist interactions between electrons
in superconductors,
and the Coulomb interaction
changes low-energy properties
of electrons moving diffusively
by disorder.~\cite{schmid,altshulerSSC1979,altshulerJETP1979}
This is how 
the scattering by nonmagnetic impurities reduces 
the transition temperature 
of s-wave superconductors.
~\cite{ovchinnikov,maekawa,takagi}
Thus, the correlation between interactions and
disorder in superconductors has been an interesting 
research subject.

Studies on correlation between the Coulomb interaction and
impurity scattering have been mainly conducted in the normal
state, and 
physical quantities such as specific heat and
conductivity have been calculated not only in
the three-dimensional case~\cite{altshulerSSC1979,altshulerJETP1979},
but also in the two-dimensional system.~\cite{altshulerPRL1980,fukuyama,abrahams}
The deviation of physical properties from those of a
Fermi liquid is
caused by the suppression
of low-energy electronic states owing to
the Coulomb interaction enhanced by diffuson.
This suppression of the density of states (DOS)
near the Fermi level is known as the Altshuler-Aronov effect.
Not only the screened Coulomb interaction but 
the superconducting fluctuation is also enhanced 
by the diffusive motion of electrons, and
this effect results in the suppression of the DOS
above the superconducting transition temperature.~\cite{abrahams1970,dicastro}

There have been several measurements 
on the DOS
both in the ultrathin film~\cite{sacepe2008,carbillet2016}
(whose thickness is comparable to the coherence length) and
the three-dimensional system.~\cite{chockalingam,chand,kamlapure}
These studies mainly focus on 
the physical properties near the superconductor-insulator
transition, especially
the variation of the size of the superconducting gap
and its spatial distribution
when the disorder is increased.
For this reason, although
the DOS exhibits 
an energy dependence 
similar to that of the Altshuler-Aronov effect
both above and below 
the superconducting transition temperature,
this energy dependence
is treated as a uniform background.
Therefore, the dependence of the DOS
on energy in the superconducting state
has not been precisely investigated.

In this study, we calculated the correction to the DOS
in the superconducting state of a three-dimensional system.
We considered the weakly localized regime in which
the expansion parameter of the perturbation
is $1/k_F l$ ($k_F$ and $l$ being the Fermi wave number
and the mean free path, respectively).
We also assume the dirty limit 
($\Delta\tau\ll 1$. $\Delta$ and $\tau$ being the superconducting gap
and the relaxation time, respectively).
In the calculation the Coulomb interaction
is included consistently with the superconducting correlation.

Although the Altshuler-Aronov effect in the superconducting
state has been studied with use of
the Coulomb interaction and diffuson of the normal state,~\cite{browne,rabatin}
the Coulomb interaction
and the effect of disorder 
are modified in the superconducting state.
The density fluctuation
couples to the fluctuation of the phase of the superconducting
order parameter.~\cite{anderson58}
In addition, because there is an energy gap in the superconducting
state,
the diffusive motion of quasiparticles is modified
and the calculation in the normal state does not hold at
low energy.
Therefore, in the vicinity of the energy gap,
the correction to the DOS also differs from
that of the normal state.

This paper is organized as follows.
In Sect. 2,
the expression for DOS is derived,
after discussing the model and the approximations 
required to calculate the correction to the DOS.
In Sect. 3, after discussing the temperature dependence and
diffuson in the superconducting state, the
results of numerical calculations at absolute zero are
presented.
In Sect. 4, a short summary is provided along with a
discussion of the effects that are not included in this paper.

\section{Formulation}

The Hamiltonian is given by 
\begin{equation}
  \begin{split}
  {\cal H}=&
  \sum_{k,\sigma} \xi_k c^{\dagger}_{k,\sigma}c_{k,\sigma}
  +\sum_q \omega_q b^{\dagger}b_q
  +\frac{g_{ph}}{\sqrt{N^3}}\sum_{k,q,\sigma}(b_q+b^{\dagger}_{-q})
  c^{\dagger}_{k+q,\sigma}c_{k,\sigma}
  +\frac{1}{\sqrt{N^3}}\sum_{k,k',\sigma}u_{k-k'}
  c^{\dagger}_{k,\sigma}c_{k',\sigma} \\
&  +\frac{1}{2N^3}\sum_{k,k',q,\sigma,\sigma'}v_q
  c^{\dagger}_{k,\sigma}c_{k+q,\sigma}
  c^{\dagger}_{k',\sigma'}c_{k'-q,\sigma'}.
  \end{split}
  \label{eq:hamiltonian}
\end{equation}
$\xi_k$ and $\omega_q$ are the dispersions of electrons
and phonons, respectively.
The third and fourth terms represent
the interaction between electrons and phonons and
the effect of impurity scattering, respectively.
We assume that $\omega_q$ does not depend on $q$
and that it takes a constant value $\omega_q=\omega_E$.
The fifth term represents the Coulomb interaction
between electrons and 
$v_q=4\pi e^2/q^2$. $N^3$ is the number of sites.
We consider the three-dimensional system,
and $k$ and $q$ are wave number vectors in this space.
We set $\hbar=1$ in this paper.

The correction to the DOS is given by
\begin{equation}
\rho'(\epsilon)=  \frac{-1}{\pi}{\rm Im}\frac{1}{N^3}\sum_{\mib k}
     {\rm Tr}  [\hat{G}_k\hat{G}'_{k}\hat{G}_k]_{i\epsilon_n\to
       \epsilon+i0^+}.
\end{equation}
Hereafter, we use the notation $k=({\mib k},\epsilon_n)$,
where 
${\mib k}$ is a wave number vector in the three dimensional space
and
$\epsilon_n=\pi T (2n-1)$ is the Matsubara frequency with
$T$ the temperature.
The term
${\rm Im}$ indicates the imaginary part, and
$i\epsilon_n\to\epsilon+i0^+$ means the analytic continuation, 
with $0^+$ an infinitesimal positive quantity ($i=\sqrt{-1}$).
$\hat{G}_k$ is the Green function of electrons and
includes the effects of the impurity scattering
and the electron-phonon interaction with 
Born and mean-field approximations,~\cite{abrikosov59}
respectively, 
\begin{equation}
  \hat{G}_k=
\frac{1}{(i\tilde{\epsilon}_n)^2-\xi_{\mib k}^2-\tilde{\Delta}^2}
  \begin{pmatrix}
    i\tilde{\epsilon}_n+\xi_{\mib k} & \tilde{\Delta} \\
    \tilde{\Delta} & i\tilde{\epsilon}_n-\xi_{\mib k}
  \end{pmatrix}.
\end{equation}
Here,
$\tilde{\epsilon}_n$ and $\tilde{\Delta}$ are
determined by the following equation:
\begin{equation}
  (i\epsilon_n-i\tilde{\epsilon}_n)\hat{\tau}_3
  +(\tilde{\Delta}-\Delta)\hat{\tau}_1=
  \frac{n_i u^2}{N^3}\sum_{\mib k}
  \hat{\tau}_3\hat{G}_k\hat{\tau}_3
\end{equation}
where $\hat{\tau}_3=\left(\begin{smallmatrix}
  1 & 0 \\ 0 & -1\end{smallmatrix}\right)$
  and
$\hat{\tau}_1=\left(\begin{smallmatrix}
  0 & 1 \\ 1 & 0\end{smallmatrix}\right)$.  
  ($n_i$ and $u$ represent the concentration of impurities and
the magnitude of the impurity potential, respectively.)
$\Delta$ is the superconducting gap determined by the gap equation,
\begin{equation}
  \Delta\hat{\tau_1}=
  \frac{g_{ph}^2}{\omega_E}\frac{2T}{N^3}\sum_k
  \hat{\tau}_3\hat{G}_k\hat{\tau}_3.
\end{equation}

\begin{figure}
\includegraphics[width=11.5cm]{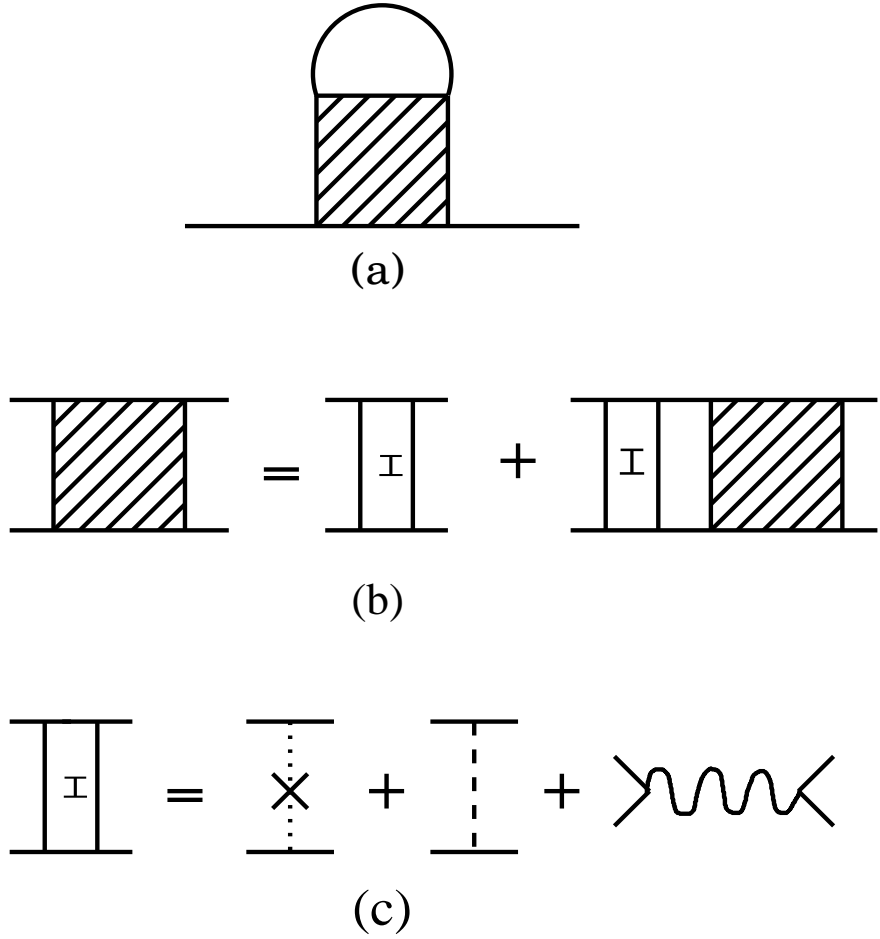}
\caption{\label{fig:1}
  (a) The diagrammatic representation of the correction
  to the DOS. The solid line indicates the propagator
  of electrons $\hat{G}_k$, and the shaded square
  includes the effects of interactions.
  (b) The interaction effect is obtained by solving this equation.
  The square with ``$I$'' included indicates the irreducible part.
  (c) The irreducible part. The dotted line with a cross
  represents the scattering by impurities. The dashed line means
  the electron-phonon interaction. The wavy line represents
the Coulomb interaction.}
\end{figure}

  The effects of interactions beyond the mean-field
  approximation are included in $\hat{G}'_{k}$,
  and its diagrammatic representation is shown in Fig.~\ref{fig:1}.
The three interaction terms in Fig. 1(c) combined with
the equation represented by Fig. 1(b) give 
the physical effects that are predominant at low energy.
The first term (the scattering by impurities) induces
the diffusive motion of electrons,
the second term (the interaction of electrons
with phonons) results in the superconducting fluctuation,
and the third term gives
the screened Coulomb interaction.

We obtain $\hat{G}'_{k}$ as follows.
The components of $\hat{G}'_{k}$ are given by
\begin{equation}
(\hat{G}'_{k})_{jj'}=\frac{-2T}{N^3}\sum_q
  \tilde{\gamma}^{ij,i'j'}_{k,k-q}(\hat{G}_{k-q})_{ii'}
\end{equation}
in which $i,j,i',j'$ are indices specifying rows and columns
of $2\times 2$ matrices; hereafter
the summation is taken over repeated indices.
$\tilde{\gamma}^{ij,i'j'}_{k,k-q}$ is
given by 
\begin{equation}
  \tilde{\gamma}^{ij,i'j'}_{k,k-q}=
  \left[\delta_{i,s}\delta_{j,t}+
    n_i u^2M^{ij,lm}_{\epsilon_n,\epsilon_n-\omega_l}
    (\hat{\tau}_3)_{sl}(\hat{\tau}_3)_{mt}\right]
  \gamma^{st,s't'}_{q}
  \left[\delta_{i',s'}\delta_{j',t'}+
(\hat{\tau}_3)_{l's'}(\hat{\tau}_3)_{t'm'}
    n_i u^2M^{l'm',i'j'}_{\epsilon_n,\epsilon_n-\omega_l}\right],
\end{equation}
where $\delta_{i,s}$ is Kronecker's delta function. 
$\gamma^{ij,i'j'}_{q}$ and 
$M^{ij,i'j'}_{\epsilon_n,\epsilon_n-\omega_l}$
are given by the following equations.
\begin{equation}
  \begin{split}
  \gamma^{ij,i'j'}_q=&
\left[\frac{g_{ph}^2}{\omega_E}
(\hat{\tau}_3)_{i'i}(\hat{\tau}_3)_{jj'}
  +\frac{v_q}{2}
(\hat{\tau}_3)_{ij}(\hat{\tau}_3)_{i'j'}\right] \\
&+\left[\frac{g_{ph}^2}{\omega_E}
(\hat{\tau}_3)_{li}(\hat{\tau}_3)_{jm}
  +\frac{v_q}{2}
  (\hat{\tau}_3)_{ij}(\hat{\tau}_3)_{lm}\right]
2T\sum_{\epsilon_n}
M^{lm,l'm'}_{\epsilon_n,\epsilon_n-\omega_l}
\gamma^{l'm',i'j'}_q
  \end{split}
  \label{eq:eqforgamma}
\end{equation}
and
\begin{equation}
  M^{ij,i'j'}_{\epsilon_n,\epsilon_n-\omega_l}=
  \frac{1}{N^3}\sum_{\mib k}
  (\hat{G}_{k})_{jj'}(\hat{G}_{k-q})_{i'i}
+  \frac{n_i u^2}{N^3}\sum_{\mib k}
  (\hat{G}_{k})_{jm}(\hat{G}_{k-q})_{li}
(\hat{\tau}_3)_{l'l}(\hat{\tau}_3)_{mm'}
M^{l'm',i'j'}_{\epsilon_n,\epsilon_n-\omega_l}.
\label{eq:eqforM}
\end{equation}
These equations are solved by introducing $4\times 4$
matrices such as
\begin{equation}
\hat{M}:=  
  \begin{pmatrix}
    M^{11,11}
    & M^{11,22} & M^{11,12} & M^{11,21} \\
    M^{22,11} & M^{22,22} & M^{22,12} & M^{22,21} \\
    M^{12,11} & M^{12,22} & M^{12,12} & M^{12,21} \\
    M^{21,11} & M^{21,22} & M^{21,12} & M^{21,21} \\
  \end{pmatrix}.
  \label{eq:4by4}
\end{equation}
Then, for example,
\begin{equation}
(\hat{\tau}_3)_{i'i}(\hat{\tau}_3)_{jj'}
 =
  \begin{pmatrix}
    1 & 0 & 0 & 0 \\
    0 & 1 & 0 & 0 \\
    0 & 0 & -1 & 0 \\
    0 & 0 & 0 & -1 \\
  \end{pmatrix}
\end{equation}
and
\begin{equation}
(\hat{\tau}_3)_{ij}(\hat{\tau}_3)_{i'j'}
 =
  \begin{pmatrix}
    1 & -1 & 0 & 0 \\
    -1 & 1 & 0 & 0 \\
    0 & 0 & 0 & 0 \\
    0 & 0 & 0 & 0 \\
  \end{pmatrix}.
\end{equation}

By solving Eq. (\ref{eq:eqforM}),
the $4\times 4$ matrix corresponding to
$2T\sum_{\epsilon_n}M^{ij,i'j'}_{\epsilon_n,\epsilon_n-\omega_l}$
is written as follows:
\begin{equation}
    \frac{\pi\rho_0}{2}
 \begin{pmatrix}
  (\chi_3+\chi_0)
  \hat{\tau_0}/2
  -(\chi_3-\chi_0)
  \hat{\tau_1}/2
  &
  \chi'(\hat{\tau_0}-\hat{\tau_1})
  \\
  \chi'(\hat{\tau_0}-\hat{\tau_1})
  &
    (\chi_2+\chi_1)
  \hat{\tau_0}/2
  -(\chi_2-\chi_1)
  \hat{\tau_1}/2
 \end{pmatrix}.
 \label{eq:matrixchi}
\end{equation}
Here, $\hat{\tau}_0=
\left(\begin{smallmatrix} 1 & 0 \\ 0 & 1\end{smallmatrix}\right)$,
and $\rho_0=m k_F/\pi^2$ is
  the noninteracting density of states at the Fermi level.
  \begin{equation}
    \chi_i=
    2T\sum_{\epsilon_n}
      \frac{X_{\epsilon_n+\omega_l,\epsilon_n}/\alpha}
           {1-2X_{\epsilon_n+\omega_l,\epsilon_n}}
    (h_i+ g_{\epsilon_n+\omega_l}g_{\epsilon_n}
    +h_i' f_{\epsilon_n+\omega_l}f_{\epsilon_n})
    -\frac{2}{\pi}(\delta_{i,3}+\delta_{i,0})
    \label{eq:eqforchi0}    
  \end{equation}
  (the second term is necessary when the integration over $\xi_{\mib k}$
  is performed before the summation over $\epsilon_n$~\cite{AGD}),
  and
  \begin{equation}
    \chi'=
    T\sum_{\epsilon_n}
      \frac{X_{\epsilon_n+\omega_l,\epsilon_n}/\alpha}
           {1-2X_{\epsilon_n+\omega_l,\epsilon_n}}
    (g_{\epsilon_n+\omega_l}f_{\epsilon_n}
           -f_{\epsilon_n+\omega_l}g_{\epsilon_n}).
  \end{equation}  
  $g_{\epsilon_n}=-i\epsilon_n/\zeta_{\epsilon_n}$,
  $f_{\epsilon_n}=-\Delta/\zeta_{\epsilon_n}$,
  $\zeta_{\epsilon_n}=\sqrt{\epsilon_n^2+\Delta^2}$,
  $\alpha:=n_i u^2m k_F/2\pi$,
  \begin{equation}
    h_3=h_0=h_0'=h_1'=1,
    \end{equation}
  and
  \begin{equation}
    h_3'=h_2=h_2'=h_1=-1.
  \end{equation}
  $\alpha$ is related to the relaxation time by the impurity scattering: 
  $\tau=1/2\alpha=1/\pi\rho_0n_iu^2$. 
\begin{equation}
  X_{\epsilon_n,\epsilon_{n'}}:=
  \int_{FS}\frac{2\alpha+\zeta_{\epsilon_n}
    +\zeta_{\epsilon_{n'}}}
      {(2\alpha+\zeta_{\epsilon_n}+\zeta_{\epsilon_{n'}})^2+(v_k\cdot q)^2}
      =
      \frac{2\alpha}{v_F q}{\rm arctan}
      \left(\frac{v_F q}{2\alpha+\zeta_{\epsilon_n}+\zeta_{\epsilon_{n'}}}
      \right).
\end{equation}
($\int_{FS}$ indicates the integration over the Fermi surface.)
In the case of a dirty limit ($v_F q/2\alpha\ll 1$,
$(\zeta_{\epsilon_n}+\zeta_{\epsilon_{n'}})/2\alpha\ll 1$)
\begin{equation}
X_{\epsilon_n,\epsilon_{n'}}
  \simeq
  \frac{2\alpha-(D_{\alpha}q^2+\zeta_{\epsilon_n}
    +\zeta_{\epsilon_{n'}})}
       {4\alpha}
       \label{eq:approXdiff}
\end{equation}
with the diffusion constant $D_{\alpha}=v_F^2\tau/3$
($v_F$ is the Fermi velocity).

The indices $i$ of $\chi_i$ correspond to
those of Pauli matrices ($\hat{\tau}_i$).
Using Eq. (\ref{eq:matrixchi}),
\begin{equation}
\left(    \frac{\pi\rho_0}{2}\right)^{-1}
  (2T\sum_{\epsilon_n}M^{ij,i'j'}_{\epsilon_n,\epsilon_n-\omega_l})
  (\hat{\tau}_{0,1})_{ii'}
  =\chi_{0,1}(\hat{\tau}_{0,1})_{jj'},
\end{equation}
and
\begin{equation}
\left(    \frac{\pi\rho_0}{2}\right)^{-1}  
  (2T\sum_{\epsilon_n}M^{ij,i'j'}_{\epsilon_n,\epsilon_n-\omega_l})
  (\hat{\tau}_{3}+i\hat{\tau}_{2})_{ii'}
=(\chi_{3}+2\chi')(\hat{\tau}_{3})_{jj'}
+(\chi_{2}+2\chi')(i\hat{\tau}_{2})_{jj'}. 
\end{equation}
$\hat{\tau}_2=\left(\begin{smallmatrix}
  0 & -i \\ i & 0\end{smallmatrix}\right)$.
These equations indicate that
the density fluctuation ($\hat{\tau}_3$)
couples to the phase fluctuation ($\hat{\tau}_2$)
in the presence of a finite value of the superconducting gap
(the mixing term $\chi'$ vanishes when $\Delta=0$),
and the amplitude fluctuation ($\hat{\tau}_1$) decouples from
other modes in the presence of a particle-hole symmetry. 

Then, the solution for Eq. (\ref{eq:eqforgamma}) is written
in the $4\times 4$ matrix form as follows:
\begin{equation}
  \hat{\gamma}_q=\left(    \frac{\pi\rho_0}{2}\right)^{-1}
  \begin{pmatrix}
    \Gamma_3(q)(\hat{\tau}_0-\hat{\tau}_1)
    +\Gamma_0(q)(\hat{\tau}_0+\hat{\tau}_1)
    & \Gamma'(q)(\hat{\tau}_0-\hat{\tau}_1) \\
    \Gamma'(q)(\hat{\tau}_0-\hat{\tau}_1) &
     \Gamma_2(q)(\hat{\tau}_0-\hat{\tau}_1)
     +\Gamma_1(q)(\hat{\tau}_0+\hat{\tau}_1)
  \end{pmatrix}.
\end{equation}
Here, 
  \begin{equation}
    \Gamma_3(q)=\frac{(p+c_q) (1/p+\chi_2)/2}
          {(1/p+\chi_2)[1-(p+c_q)\chi_3]
            +4(p+c_q)(\chi')^2},
\label{eq:gamma30}
  \end{equation}
  \begin{equation}
    \Gamma_2(q)=\frac{-[1-(p+c_q) \chi_3]/2}
          {(1/p+\chi_2)[1-(p+c_q)\chi_3]
            +4(p+c_q)(\chi')^2},
\label{eq:gamma20}          
  \end{equation}  
  \begin{equation}
    \Gamma'(q)=\frac{(p+c_q) \chi'}
          {(1/p+\chi_2)[1-(p+c_q)\chi_3]
            +4(p+c_q)(\chi')^2},
\label{eq:gammad0}          
  \end{equation}
  \begin{equation}
    \Gamma_0(q)=\frac{p/2}
          {1-p\chi_0},
\label{eq:gamma00}          
  \end{equation}
  and
  \begin{equation}
    \Gamma_1(q)=\frac{-1/2}
          {1/p+\chi_1}.
\label{eq:gamma10}          
  \end{equation}
  Here, 
  $p:=m k_Fg_{ph}^2/2\pi\omega_E$ indicates the coupling
  constant between electrons and phonons
  and $c_q:=m k_F v_q/2\pi$.

Using the above results, 
the correction to the DOS is written as follows.
  \begin{equation}
    \begin{split}
&    \frac{-1}{\pi N^3}\sum_k{\rm Tr}
         [\hat{G}_k\hat{G}'_k\hat{G}_k]\simeq
         \rho_0\frac{
           3\sqrt{3\tau}}{2\pi(k_Fl)^2}
      2T\sum_{\omega_l}\int dx \sqrt{x} \\
      &\times
        \frac{\Gamma_i(q)(h_i+g_{\epsilon_n}g_{\epsilon_n-\omega_l}
      +h_i'f_{\epsilon_n}f_{\epsilon_n-\omega_l})
+2\Gamma'(q)(f_{\epsilon_n}g_{\epsilon_n-\omega_l}
-g_{\epsilon_n}f_{\epsilon_n-\omega_l})}
 { (x+\zeta_{\epsilon_n}+\zeta_{\epsilon_n-\omega_l})^2
        }          g_{\epsilon_n}.
         \end{split}
    \label{eq:scdoscorr}
  \end{equation}
  ($x=D_{\alpha}q^2$.)
  Here we use the approximate expression Eq. (\ref{eq:approXdiff}),
  and introduce the upper limits of
  $|\omega_l|$ and $D_{\alpha}q^2$ (which
  are on the order of $2\alpha$ and will be specified when
  the numerical calculation is performed in Sect. 3).
(The high energy parts from $|\omega_l|/2\alpha\gg 1$ or
  $v_F q/2\alpha\gg 1$ are assumed to be included in
  the parameters of the electronic states.
  In fact, $1/(1-2X_{\epsilon_n,\epsilon_n-\omega_l})\simeq 1$
  in this range, and the correction term is reduced to
  the usual Fock term because 
  the diffuson propagator is absent.)

\subsection{Normal state}
In this subsection, we show that
the expressions previously studied in the normal state 
~\cite{altshulerSSC1979,altshulerJETP1979,abrahams1970,dicastro}
are obtained by setting $\Delta=0$ in the above expressions.
For $\Delta=0$ and after analytic continuation
($i\omega_l\to \omega+i0^+$) $\chi_i$ ($i=0,1,2,3$) and
$\chi'$ are written as follows.
\begin{equation}
  \chi_3=\chi_0=
  \frac{2}{\pi}
  \frac{-D_{\alpha}q^2}{D_{\alpha}q^2-i\omega},
\end{equation}
\begin{equation}
  \frac{1}{p}+  \chi_2=
  \frac{1}{p}+  \chi_1=
  \frac{2}{\pi}\int d\epsilon
  \left[
    \frac{{\rm tanh}(\epsilon/2T_c)}{2\epsilon}
    +\frac{-{\rm tanh}(\epsilon/2T)}{2\epsilon+\omega+iD_{\alpha}q^2}
    \right]
  \simeq
  \frac{2}{\pi}
  \left[
    {\rm ln}\left(\frac{T}{T_c}\right)
    +\frac{\pi}{8T}(D_{\alpha}q^2-i\omega)
    \right]
  \label{eq:nmchi2chi1}
\end{equation}
($T_c$ is the superconducting transition temperature) and $\chi'=0$.
Then, $\Gamma_i(q)$ and $\Gamma'(q)$ are given by
\begin{equation} 
  \Gamma_3(q)=\frac{(p+c_q)(1/p+\chi_2)/2}
        {(1/p+\chi_2)[1-(p+c_q)\chi_3]}
        \simeq \frac{-1/2}{\chi_3},
        \label{eq:nmgamma3}
\end{equation}
\begin{equation}
  \Gamma_2(q)=\frac{-1/2}{1/p+\chi_2}=\Gamma_1(q), 
\end{equation}
\begin{equation}
  \Gamma_0(q)=\frac{p/2}{1-p\chi_3},
\end{equation}
and $\Gamma'(q)=0$.

The correction to the DOS in the normal state
is given by the following equation:
  \begin{equation}
    \begin{split}
&\rho'(\epsilon)
      =
\rho'_{sf}(\epsilon)      +\rho'_{cl}(\epsilon)      
    \end{split}
    \label{eq:nmdoscorr}
  \end{equation}
with
  \begin{equation}
    \begin{split}
&\rho'_{sf}(\epsilon)
\simeq 
\rho_0
   \frac{12\sqrt{3\tau}}{(2\pi k_F l)^2}
     \int d\omega\int d x\sqrt{x}
       {\rm Im}\left\{
            \frac
                {2i{\rm coth}(\frac{\omega}{2T})
                    {\rm Im}[\Gamma_2(q)]
+{\rm tanh}(\frac{\epsilon-\omega}{2T})\Gamma_2(q)}
                 {[x-i(2\epsilon-\omega)]^2}                     
                 \right\}
    \end{split}
    \label{eq:nmsfdoscorr}
  \end{equation}
and
  \begin{equation}
    \begin{split}
&\rho'_{cl}(\epsilon)
\simeq 
\rho_0
   \frac{6\sqrt{3\tau}}{(2\pi k_F l)^2}
     \int d\omega\int d x\sqrt{x}           {\rm Im}\left\{
\frac{{\rm tanh}(\frac{\epsilon-\omega}{2T})[\Gamma_3(q)+\Gamma_0(q)]}{(x-i\omega)^2}
                 \right\}.
    \end{split}
    \label{eq:nmcldoscorr}
  \end{equation}
  $\rho'_{sf}(\epsilon)$   and $\rho'_{cl}(\epsilon)$ include
  the effects of the superconducting fluctuation above
  $T_c$~\cite{abrahams1970,dicastro}
  and the screened Coulomb interaction
  enhanced by
  diffuson,~\cite{altshulerSSC1979,altshulerJETP1979}
  respectively.

\section{Results}

\subsection{The temperature dependence of the correction
  to the density of states}

In this subsection, we show that the temperature dependence of
the correction to DOS is small at low temperature $T\ll\Delta$.

After analytic continuation, Eq. (\ref{eq:eqforchi0}) is written
as follows.
  \begin{equation}
    \chi_i=
    \int\frac{d\epsilon}{2\pi i}
    \left[
      {\rm tanh}\left(\frac{\epsilon}{2T}\right)
      (\kappa^i_{++}-\kappa^i_{+-})
      +
            {\rm tanh}\left(\frac{\epsilon+\omega}{2T}\right)
      (\kappa^i_{+-}-\kappa^i_{--})
            \right]
    -\frac{2}{\pi}(\delta_{i,3}+\delta_{i,0})
    \label{eq:eqforchi}
  \end{equation}
    with
  \begin{equation}
      \kappa^i_{s s'}=
      \frac{X^{s s'}_{\epsilon+\omega,\epsilon}/\alpha}
           {1-2X^{s s'}_{\epsilon+\omega,\epsilon}}
    (h_i+ g^{s}_{\epsilon+\omega}g^{s'}_{\epsilon}
           +h_i' f^{s}_{\epsilon+\omega}f^{s'}_{\epsilon}).
  \end{equation}
  $\chi'$ is obtained by
  replacing $\kappa^i_{s s'}$ in Eq. (\ref{eq:eqforchi})
  with $i\ne 3,0$ by 
  \begin{equation}
  \kappa'_{s s'}=
  \frac{X^{s s'}_{\epsilon+\omega,\epsilon}/\alpha}
       {1-2X^{s s'}_{\epsilon+\omega,\epsilon}}
    (g^{s}_{\epsilon+\omega}f^{s'}_{\epsilon}
       -f^{s}_{\epsilon+\omega}g^{s'}_{\epsilon})/2.
\end{equation}
  Here, 
  $s,s'=+$ (retarded) or $-$ (advanced),
  $g^{s}_{\epsilon}=-\epsilon/\zeta^s_{\epsilon}$, and
  $f^{s}_{\epsilon}=-\Delta/\zeta^s_{\epsilon}$ with
$\zeta^{\pm}_{\epsilon}=
\sqrt{\Delta^2-\epsilon^2}\theta(\Delta-|\epsilon|)
-i{\rm sgn}(\pm\epsilon)\sqrt{\epsilon^2-\Delta^2}\theta(|\epsilon|-\Delta)$ [$\theta(\cdot)$ is a step function]. 
\begin{equation}
  X^{ss'}_{\epsilon,\epsilon'}=
  \int_{FS}\frac{2\alpha+\zeta^s_{\epsilon}+\zeta^{s'}_{\epsilon'}}
      {(2\alpha+\zeta^s_{\epsilon}+\zeta^{s'}_{\epsilon'})^2+(v_k\cdot q)^2}
  \simeq
  \frac{2\alpha-(D_{\alpha}q^2+\zeta^s_{\epsilon}+\zeta^{s'}_{\epsilon'})}
       {4\alpha}.
\end{equation}
From Eq. (\ref{eq:eqforchi}),
\begin{equation}
  {\rm Im}\chi_i=\int\frac{d\epsilon}{2\pi}
  \left[{\rm tanh}\left(\frac{\epsilon+\omega}{2T}\right)
    -{\rm tanh}\left(\frac{\epsilon}{2T}\right)\right]
       {\rm Re}(\kappa^i_{++}-\kappa^i_{+-}).
\end{equation}
${\rm Re}(\kappa^i_{++}-\kappa^i_{+-})$ takes finite values
only for $|\epsilon+\omega|>\Delta$ and $|\epsilon|>\Delta$.
Then,
${\rm Im}\chi_i$ is exponentially small
for $|\omega|<2\Delta$ except for $T\simeq T_C$,
and is negligible in this region.

We consider the correction to the DOS for $|\epsilon|<\Delta$
and $|\epsilon|>\Delta$  separately in the following.
First, we consider the case of $|\epsilon|<\Delta$.
After performing the analytic continuation
of Eq. (\ref{eq:scdoscorr}),
the imaginary part is written as follows.
\begin{equation}
    \begin{split}
&\rho'(\epsilon)
\simeq
   \frac{\rho_0\epsilon}{\sqrt{\Delta^2-\epsilon^2}}
   \frac{-6\sqrt{3\tau}}{(2\pi k_F l)^2}
   \int d\omega\int d x\sqrt{x}
   \left[{\rm coth}\left(\frac{\omega}{2T}\right)+
     {\rm tanh}\left(\frac{\epsilon-\omega}{2T}\right)\right] \\
&\times
   {\rm Im}
\Bigl\{
\frac{{\rm Im}[\Gamma_i(q)](h_i+g_{\epsilon}g^+_{\epsilon-\omega}
       +h_i'f_{\epsilon}f^+_{\epsilon-\omega})+
2{\rm Im}[\Gamma'(q)](f_{\epsilon}g^+_{\epsilon-\omega}
            -g_{\epsilon}f^+_{\epsilon-\omega})}
     {(x+\zeta_{\epsilon}+\zeta^+_{\epsilon-\omega})^2}\Bigr\}
    \end{split}
\end{equation}
($\zeta_{\epsilon}=\sqrt{\Delta^2-\epsilon^2}$,
$g_{\epsilon}=-\epsilon/\zeta_{\epsilon}$, and
$f_{\epsilon}=-\Delta/\zeta_{\epsilon}$).
The imaginary part is finite (${\rm Im}\{\cdot\}\ne 0$) 
only for $|\epsilon-\omega|>\Delta$.
For $|\omega|<2\Delta$,
${\rm Im}\Gamma_i$ and ${\rm Im}\Gamma'$ 
are exponentially small at low temperature,
as noted above.
The factor ${\rm coth}(\omega/2T)+{\rm tanh}[(\epsilon-\omega)/2T]$
is also exponentially small for $|\epsilon|<\Delta$
and $|\omega|>2\Delta$.
  Then, the correction to the DOS
  is negligible for $|\epsilon|<\Delta$ except for $T\simeq T_C$.
  
  On the other hand, 
  for $|\epsilon|>\Delta$,
  the imaginary part of Eq. (\ref{eq:scdoscorr}) after
  the analytic continuation
    is written as follows:
\begin{equation}
    \begin{split}
&\rho'(\epsilon)
\simeq
   \frac{\rho_0|\epsilon|}{\sqrt{\epsilon^2-\Delta^2}}
   \frac{-3\sqrt{3\tau}}{(2\pi k_F l)^2}
   \int d\omega\int d x\sqrt{x} \\
&\times
   {\rm Im}
   \Bigl\{2{\rm coth}\left(\frac{\omega}{2T}\right)
\frac{{\rm Im}[\Gamma_i(q)](h_i+g^+_{\epsilon}g^+_{\epsilon-\omega}
       +h_i'f^+_{\epsilon}f^+_{\epsilon-\omega})+
2{\rm Im}[\Gamma'(q)](f^+_{\epsilon}g^+_{\epsilon-\omega}
            -g^+_{\epsilon}f^+_{\epsilon-\omega})}
          {(x+\zeta^+_{\epsilon}+\zeta^+_{\epsilon-\omega})^2}
                            \\
   &     +     {\rm tanh}\left(\frac{\epsilon-\omega}{2T}\right)
   \sum_{s=\pm}s
\frac{\Gamma_i(q)(h_i+g^+_{\epsilon}g^s_{\epsilon-\omega}
       +h_i'f^+_{\epsilon}f^s_{\epsilon-\omega})+
2\Gamma'(q)(f^+_{\epsilon}g^s_{\epsilon-\omega}
            -g^+_{\epsilon}f^s_{\epsilon-\omega})}
          {(x+\zeta^+_{\epsilon}+\zeta^s_{\epsilon-\omega})^2}\Bigr\}.
    \end{split}
    \label{eq:egddoscorr}
\end{equation}
In this equation the coefficient of ${\rm coth}(\omega/2T)$
is exponentially small for $|\omega|<2\Delta$
owing to the existence of ${\rm Im}\Gamma_i$ and ${\rm Im}\Gamma'$,
and
the coefficient of ${\rm tanh}[(\epsilon-\omega)/2T]$
vanishes for $|\epsilon-\omega|<\Delta$ (the imaginary part
is absent).
This indicates that
the dependence of $\rho'(\epsilon)$ for
$|\epsilon|>\Delta$ on temperature
is weak for $T\ll \Delta$.
This small dependence of $\rho'(\epsilon)$ on temperature
is consistent with
exponentially small values of $\rho'(\epsilon)$ for
$|\epsilon|<\Delta$ at low temperature.
Thus, we perform the numerical calculations at $T=0$
and $\epsilon>\Delta$ in Sect. 3.3.

\subsection{Diffuson in the superconducting state}

The diffuson propagator is usually 
represented by $1/(D_{\alpha}q^2-i\omega)$.
However, in the superconducting state
[Eq. (\ref{eq:egddoscorr}), $x=D_{\alpha}q^2$]
it is given by
$1/(x+\zeta^+_{\epsilon}+\zeta^{\pm}_{\epsilon-\omega})=
1/\{
x-i[{\rm sgn}(\epsilon)\sqrt{\epsilon^2-\Delta^2}
  \pm{\rm sgn}(\epsilon-\omega)\sqrt{(\epsilon-\omega)^2-\Delta^2}]\}$
for $|\epsilon|,|\epsilon-\omega|>\Delta$
(the diffusive motion of quasiparticles is effective above
the superconducting gap).
Another singularity exists at $\omega=2\epsilon$
in the case of $1/(x+\zeta^+_{\epsilon}+\zeta^{+}_{\epsilon-\omega})$
in addition to the pole at $\omega=0$ in
$1/(x+\zeta^+_{\epsilon}+\zeta^{-}_{\epsilon-\omega})$.
In this subsection, we illustrate that 
the divergence by this additional pole
is absent when the particle-number
conservation is preserved
in the integration of Eq. (\ref{eq:egddoscorr}).

By performing the analytic calculation,
\begin{equation}
  \chi_3({\mib q}={\mib 0})=
  \frac{-8\Delta^2{\rm arcsin}(\omega/2\Delta)}
       {\pi\omega\sqrt{4\Delta^2-\omega^2}}
\theta(2\Delta-\omega)
+\left[
  \frac{8\Delta^2{\rm arcosh}(\omega/2\Delta)}
       {\pi\omega\sqrt{\omega^2-4\Delta^2}}
+i\frac{-4\Delta^2}{\omega\sqrt{\omega^2-4\Delta^2}}
\right]
\theta(\omega-2\Delta)
\end{equation} 
($\omega>0$) and 
there are following relations
between $\chi_i$ ($i=0,1,2,3$) and $\chi'$ at ${\mib q}={\mib 0}$:
    $1/p+\chi_2=(\omega/2\Delta)^2\chi_3$,
    $\chi'=(-\omega/4\Delta)\chi_3$,
    $1/p+\chi_1=[(\omega/2\Delta)^2-1]\chi_3$
    and $\chi_0$=0.
    Then,
$-(1/p+\chi_2)\chi_3+4(\chi')^2=0$ at ${\mib q}={\mib 0}$.

With use of a relation
$c_q=\pi \omega_p^2\tau/2D_{\alpha}q^2\gg p$
($\omega_p$ is the plasma frequency: $\omega_p^2=4\pi n_e e^2/m$
with $n_e=k_F^3/3\pi^2$
electron density and $m$ the electron mass),
Eqs. (\ref{eq:gamma30}), (\ref{eq:gamma20}),
and (\ref{eq:gammad0}) are approximately written as follows.
\begin{equation}
  \Gamma_3(q)\simeq \frac{(1/p+\chi_2)/2}{
    -(1/p+\chi_2)\chi_3+4(\chi')^2},
  \label{eq:gamma3}
\end{equation}
\begin{equation}
  \Gamma_2(q)\simeq \frac{\chi_3/2}{
    -(1/p+\chi_2)\chi_3+4(\chi')^2},
  \label{eq:gamma2}  
\end{equation}
and
\begin{equation}
  \Gamma'(q)\simeq \frac{\chi'}{
    -(1/p+\chi_2)\chi_3+4(\chi')^2}.
  \label{eq:gammad}  
\end{equation}
This expressions show that
$\Gamma_3$,$\Gamma_2$, and $\Gamma'$ are proportional
to $1/x=1/(D_{\alpha}q^2)$
because the denominator of these quantities vanishes
at ${\mib q}={\mib 0}$.
The above relations between $\chi_i$ ($i=3,2$)
and $\chi'$ indicate that
$\Gamma_3/\Gamma' =-\omega/2\Delta$ and 
$\Gamma_2/\Gamma' =-2\Delta/\omega$ at ${\mib q}={\mib 0}$.
Then, in Eq. (\ref{eq:egddoscorr}) the term containing
$1/(x+\zeta^+_{\epsilon}+\zeta^+_{\epsilon-\omega})^2$
is proportional to the following equation:
\begin{equation}
\int d\omega  \int d x\sqrt{x}\frac{\sum_{i=3,2}\Gamma_i(q)
    (h_i+g^+_{\epsilon}g^+_{\epsilon-\omega}
    +h_i'f^+_{\epsilon}f^+_{\epsilon-\omega})
    -2\Gamma'(q)
    (g^+_{\epsilon}f^+_{\epsilon-\omega}
    -f^+_{\epsilon}g^+_{\epsilon-\omega})}
     {(x+\zeta^+_{\epsilon}+\zeta^+_{\epsilon-\omega})^2}.
     \label{eq:eq48}
\end{equation}
After the integration over $x$ with use of $\Gamma\propto 1/x$,
Eq. (\ref{eq:eq48}) is proportional to
\begin{equation}
\int d\omega
  \frac{\omega^2-4\Delta^2+(\omega^2+4\Delta^2)
    (g^+_{\epsilon}g^+_{\epsilon-\omega}
  -f^+_{\epsilon}f^+_{\epsilon-\omega})
    +4\omega\Delta
    (g^+_{\epsilon}f^+_{\epsilon-\omega}
    -f^+_{\epsilon}g^+_{\epsilon-\omega})}
       {(\zeta^+_{\epsilon}+\zeta^+_{\epsilon-\omega})^{3/2}}.
\end{equation}
Both the numerator and the denominator
of this expression vanish at $\omega=2\epsilon$, and then
the integration over $\omega$
results in a finite correction to the DOS.
Therefore, by preserving the particle-number conservation,
we obtain a finite result
even when an additional singularity exists
in the diffuson propagator in the superconducting state.
(As for the case of the pole at $\omega=0$
in Eq. (\ref{eq:egddoscorr}), 
we obtain a finite result simply
because $\Gamma_3\propto \omega^2/x$,
$\Gamma'\propto \omega/x$, and
$h_2+g^+_{\epsilon}g^-_{\epsilon-\omega}
+h_2'f^+_{\epsilon}f^-_{\epsilon-\omega}=0$ at $\omega=0$.
The relation between $\Gamma_3$, $\Gamma_2$ and
$\Gamma'$ is irrelevant in this case.)

In the case of $\Gamma_{0,1}$, the long-range part $1/x$
is absent. As for the terms containing $\Gamma_{0,1}$,
the integration over $x$ 
is proportional
to $1/(\zeta^+_{\epsilon}+\zeta^{\pm}_{\epsilon-\omega})^{1/2}$,
which results in a finite value
after the integration over $\omega$ is performed.

\subsection{Numerical calculation}

As discussed above, the dependence of
$\rho'(\epsilon)$ on temperature is weak for $T\ll T_c$,
and so we perform a numerical calculation 
at $T=0$. We consider
the superconducting gap at $T=0$ as
the unit of energy ($\Delta=1$).
$p$ is determined by the gap equation.

\begin{figure}
\includegraphics[width=11.5cm]{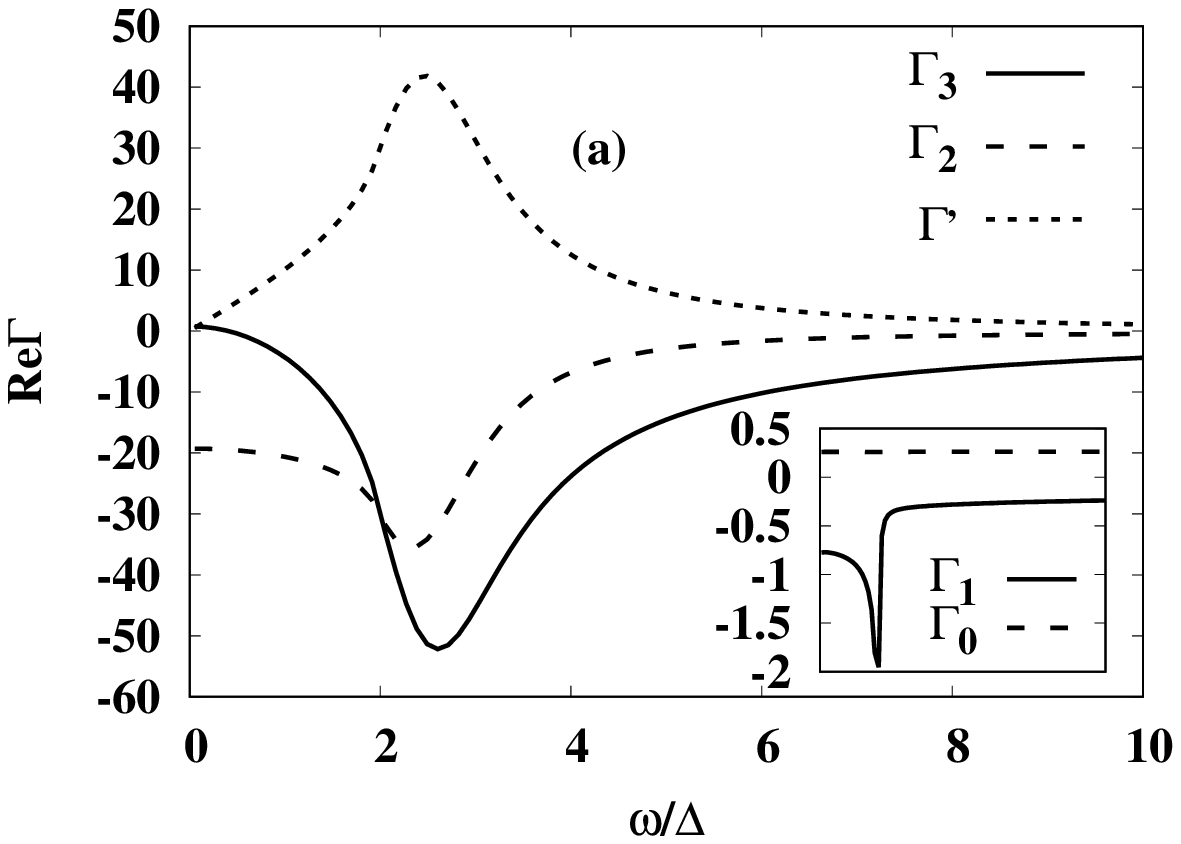}
\includegraphics[width=11.5cm]{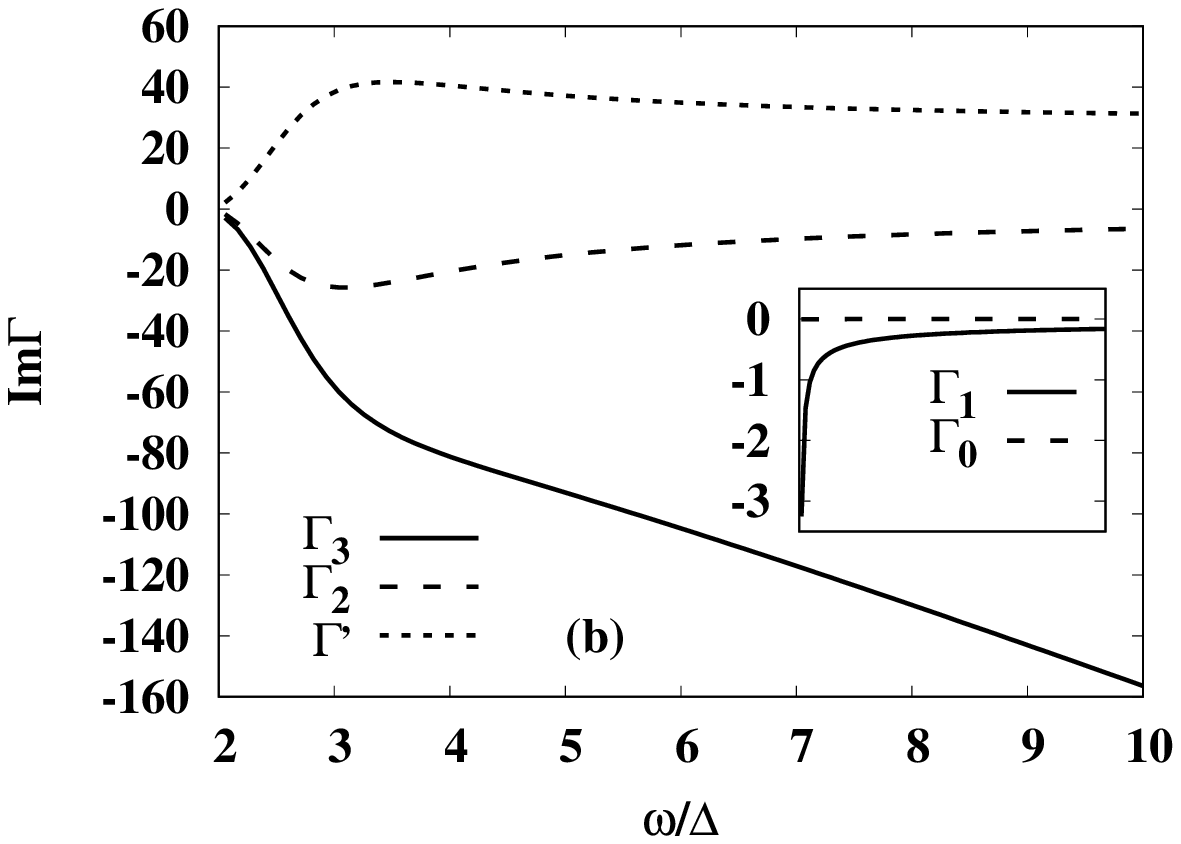}
\caption{\label{fig:2}
The dependences of $\Gamma_i$ and $\Gamma'$ on
$\omega$ at $T=0$ and $D_{\alpha}q^2/\Delta=0.055$.
(a) The real part of $\Gamma_i$ and $\Gamma'$.
(b) The imaginary part of $\Gamma_i$ and $\Gamma'$.
The ranges of $\omega/\Delta$ of the insets
are the same as those of the main graphs.
}
\end{figure}

The dependences of $\Gamma_i(q)$ and $\Gamma'(q)$
[Eqs. (\ref{eq:gamma3})$-$(\ref{eq:gammad}),
(\ref{eq:gamma00}) and (\ref{eq:gamma10})] on
$\omega$ are shown in Fig.~\ref{fig:2}.
(The value of $\alpha$ is implicitly included in
$D_{\alpha}q^2$ and the result does not depend
on $\alpha$ when the value of $D_{\alpha}q^2/\Delta$ is
fixed.)
${\rm Im}\Gamma_i$ and ${\rm Im}\Gamma'$
take finite values above $\omega>2\Delta$
owing to the finite excitation of quasiparticles
across the superconducting gap.
This leads to a peak in ${\rm Re}\Gamma$
around $\omega\simeq 2\Delta$.
For $\omega\gg\Delta$, the dependence of
$\Gamma$ on $\omega$ should become close to
that of the normal state.
The large value of ${\rm Im}\Gamma_3$ for 
$\omega\gg\Delta$ is related to
$\Gamma_3(q)\simeq (\pi/4)(1-i\omega/D_{\alpha}q^2)$
in the normal state
obtained from Eq. (\ref{eq:nmgamma3}).
The sharp peak in $\Gamma_1$ around $\omega=2\Delta$
indicates the existence of the amplitude mode.
The density and phase fluctuations
($\Gamma_3$, $\Gamma_2$, and $\Gamma'$),
however, are quantitatively predominant over $\Gamma_{1,0}$.
These large values come from the long-range part
($\propto 1/q^2$).

\begin{figure}
\includegraphics[width=11.5cm]{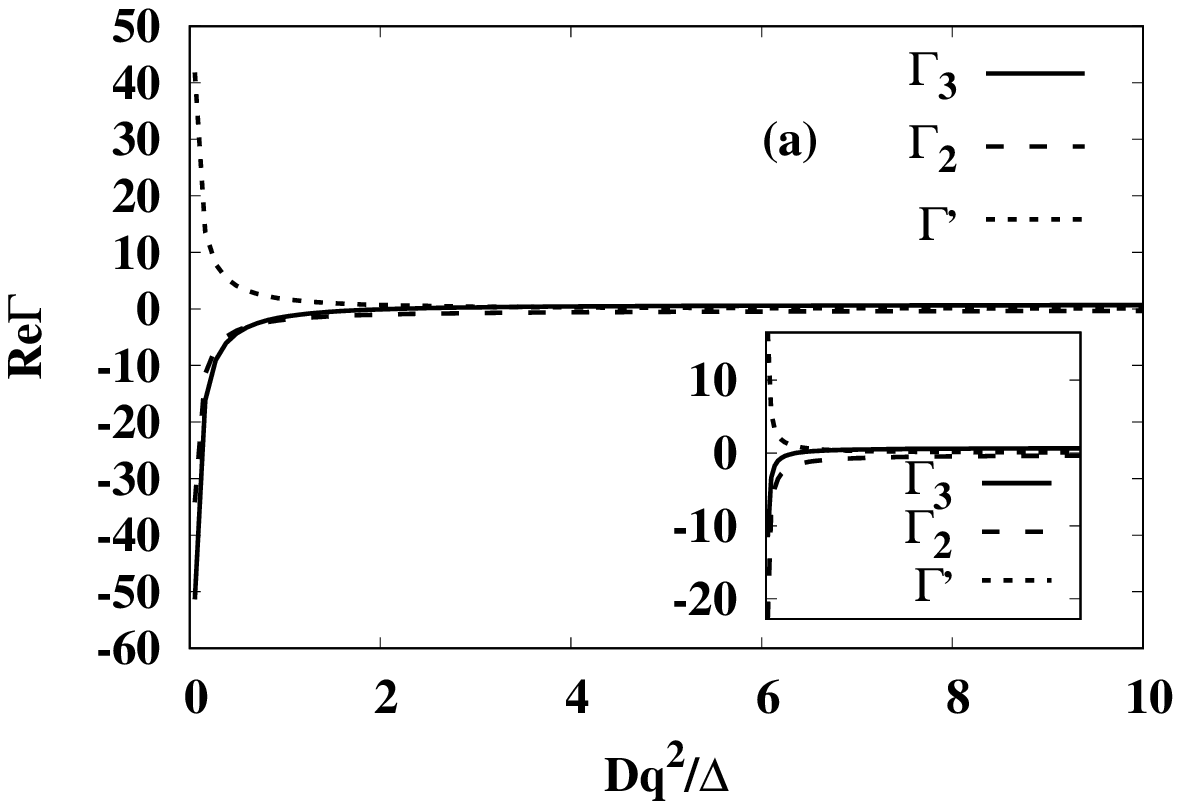}
\includegraphics[width=11.5cm]{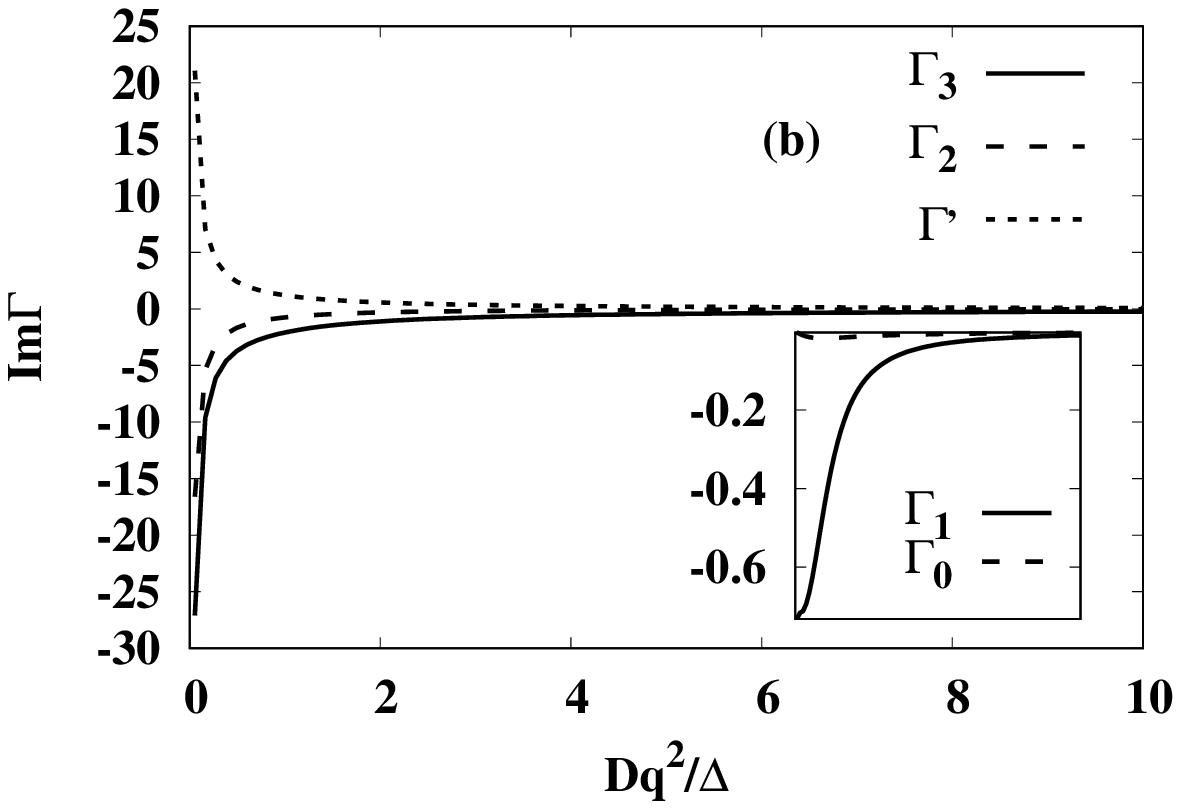}
\caption{\label{fig:3}
The dependences of $\Gamma_i$ and $\Gamma'$ on
$D_{\alpha}q^2$ at $T=0$.
(a) The real part of $\Gamma_i$ and $\Gamma'$
at $\omega/\Delta=2.5$. The inset shows the results
at $\omega/\Delta=1.48$. 
(b) The imaginary part of $\Gamma_i$ and $\Gamma'$
at $\omega/\Delta=2.5$. 
The ranges of $D_{\alpha}q^2/\Delta$ of the insets
are the same as those of the main graphs.
}
\end{figure}

The dependences of $\Gamma_i(q)$ and $\Gamma'(q)$
[Eqs. (\ref{eq:gamma3})$-$(\ref{eq:gammad}),
(\ref{eq:gamma00}) and (\ref{eq:gamma10})] on
$D_{\alpha}q^2$ are shown in Fig.~\ref{fig:3}.
$\Gamma_3$, $\Gamma_2$ and $\Gamma'$ are
proportional to $1/D_{\alpha}q^2$.
The results show that 
these three terms (the density and phase fluctuations)
are quantitatively comparable to each other.
This validates the argument about diffuson 
in the previous subsection.

Next, we calculate the correction to the DOS numerically.
From Eq. (\ref{eq:egddoscorr}), we write
the correction to DOS as follows:
\begin{equation}
  \rho'(\epsilon)=\frac{\rho_0|\epsilon|}
       {\sqrt{\epsilon^2-\Delta^2}}\delta\rho_{\epsilon}.
\end{equation}
In the case of the normal state, $\delta\rho_{\epsilon}=
\rho'(\epsilon)/\rho_0$ from Eqs. (\ref{eq:nmdoscorr})
-(\ref{eq:nmcldoscorr}).
The calculation in the superconducting state
is performed at $T=0$ as noted above.
In the case of the normal state,
the superconducting fluctuation depends on the temperature.
We fix $T=1.1T_C$ in Eq. (\ref{eq:nmchi2chi1}) and
assume $T=0$ in other terms. 
We take $|\omega|<1/\tau$ and $x=D_{\alpha}q^2<4/\tau$ as
the range of integrations 
in Eqs. (\ref{eq:nmsfdoscorr}), (\ref{eq:nmcldoscorr}) and
(\ref{eq:egddoscorr}).
The energy dependence of $\delta\rho_{\epsilon}$ is
mainly determined by the low-energy part
$|\omega|,x\ll 1/\tau$.
When we change the upper limits of $|\omega|$ and $x$,
only the magnitude of $|\delta\rho_{\epsilon}|$ is shifted.
We consider the weak-coupling case for the interaction between
electrons and phonons.
This interaction is taken to 
vanish outside the cutoff frequency ($\omega_c$),
and then $\Gamma_i(q),\Gamma'(q)\ne 0$ ($i=0,1,2$)
only for $|\epsilon|,|\epsilon-\omega|<\omega_c$
[$\Gamma_3(q)$ is finite outside this region.]
We take $\omega_c=10\Delta$ in the numerical calculation.
We specify 
the relation between $\alpha=1/2\tau$ and $k_Fl$
in Eqs. (\ref{eq:nmsfdoscorr}), (\ref{eq:nmcldoscorr}) and
(\ref{eq:egddoscorr}) 
by putting
$k_Fl/2\tau=E_F=300\Delta$ ($E_F$ is the Fermi energy).

\begin{figure}
\includegraphics[width=11.5cm]{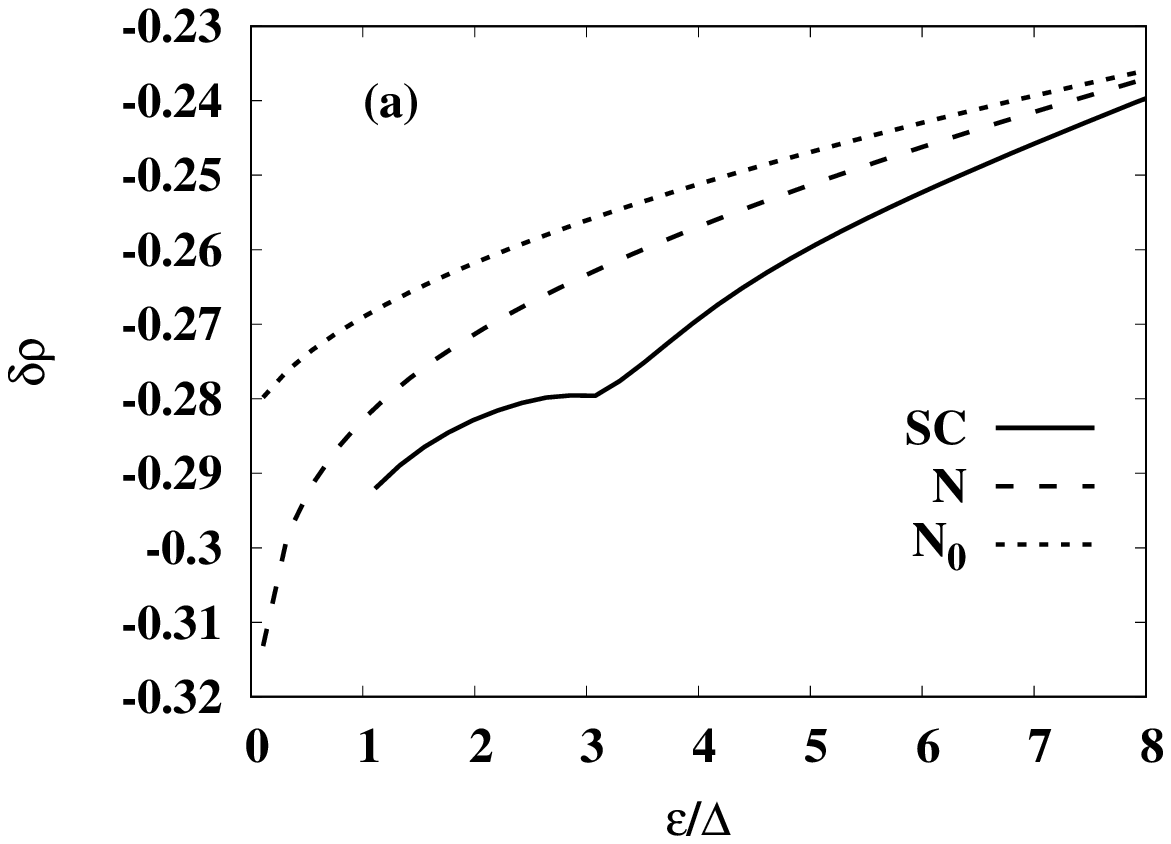}
\includegraphics[width=11.5cm]{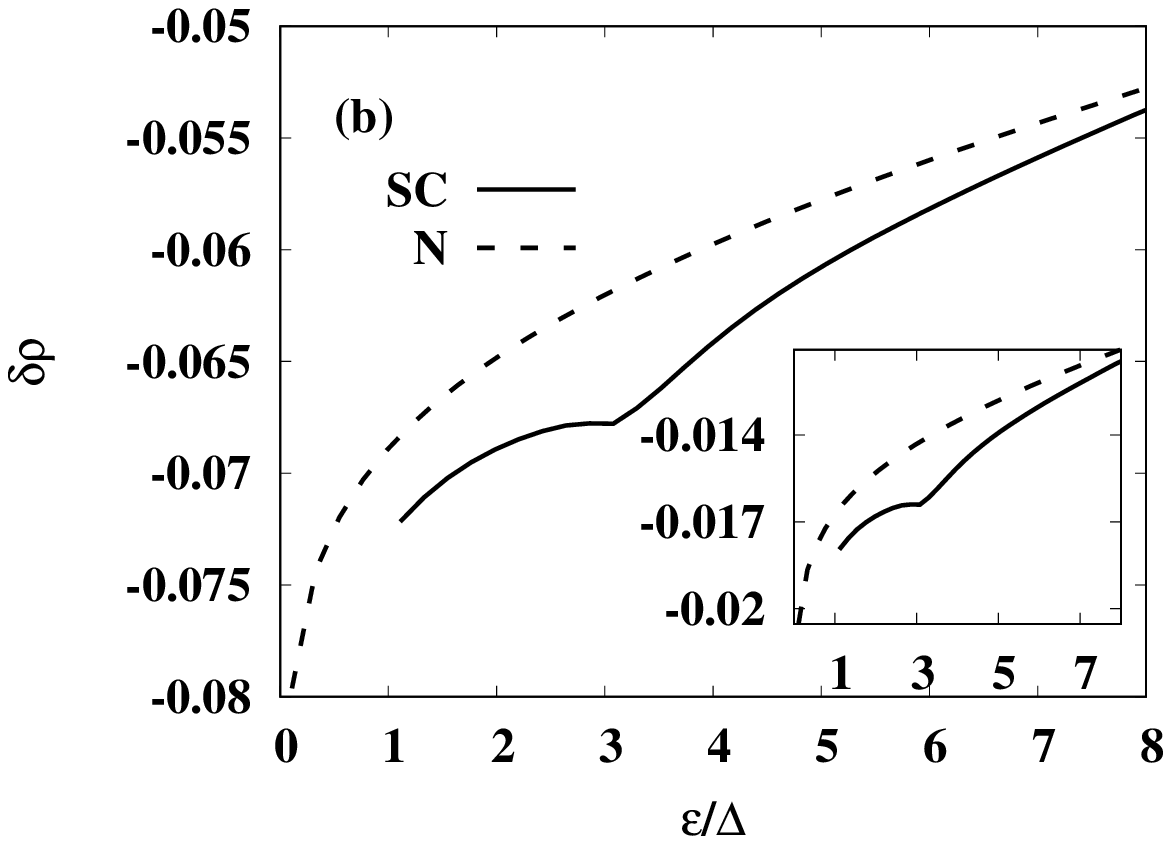}
\caption{\label{fig:4}
  The dependences of the correction to the DOS on $\epsilon$
  at $T=0$.
  (a) $\alpha/\Delta=120$ ($k_Fl=2.5$).
  (b) $\alpha/\Delta=60$ ($k_Fl=5.0$).
  The inset shows the result for
    $\alpha/\Delta=30$ ($k_Fl=10.0$).
  The meanings of ``SC'', ``N'' and ``N$_0$'' are given in the text.
}
\end{figure}

The calculated results of
the correction to the DOS are shown in Fig.~\ref{fig:4}.
``SC'' and ``N'' are the results calculated
in the superconducting state and the normal state,
respectively. 
``N$_0$'' is the calculated result with 
only the term $\Gamma_3(q)$ included in Eq. (\ref{eq:nmcldoscorr}).
The dependence of $\delta\rho_{\epsilon}$ on $\epsilon$
changes slightly with increasing $\alpha$,
and it is written as
$\delta\rho_{\epsilon}\propto \sqrt{\epsilon}$ for ``N$_0$''.
As for the dependence of the magnitude of  
$\delta\rho_{\epsilon}$ on $\alpha$,
$\delta\rho_{\epsilon}\propto 1/(k_Fl)^2$
holds
in both the superconducting and
the normal states.
This is related to 
the $\epsilon$-dependence of $\delta\rho_{\epsilon}$
because the equation
  \begin{equation}
    \begin{split}
&\rho'_{cl}(\epsilon)
\simeq 
\rho_0
   \frac{-3\sqrt{3\tau/2}}{(2 k_F l)^2}
     \int_{\epsilon}^{1/\tau} d\omega
     \frac{1}
          {\sqrt{\omega}}
          \propto
   \frac{\sqrt{\tau}}{(k_F l)^2}
 \left(-\frac{1}{\sqrt{\tau}}+\sqrt{\epsilon}\right)
    \end{split}
  \end{equation}
  is derived
  from Eq. (\ref{eq:nmcldoscorr}).

The $\epsilon$-dependences of $\delta\rho_{\epsilon}$
are not exactly written as
$\delta\rho_{\epsilon}\propto \sqrt{\epsilon}$ for ``SC'' and ``N''.
The result for ``SC'' shows that a dip structure appears
around $\epsilon=3\Delta$.
This structure is resulted from the peak in $\Gamma(q)$
around $\omega\simeq 2\Delta$.
The reason for the overall suppression in ``SC''
as compared to ``N''
is the enhancement of $\Gamma_2$
and $\Gamma'$ owing to the coupling of the phase fluctuation
to the density fluctuation. 
The result for ``N'' shows that the superconducting fluctuation 
suppresses the DOS at low energy.
The $\delta\rho_{\epsilon}$ values of ``SC'' and ``N'' approach
that of ``N$_0$'' at high energy owing to the weakening
of the superconducting correlation for $\epsilon\gg\Delta$.

\begin{figure}
\includegraphics[width=11.5cm]{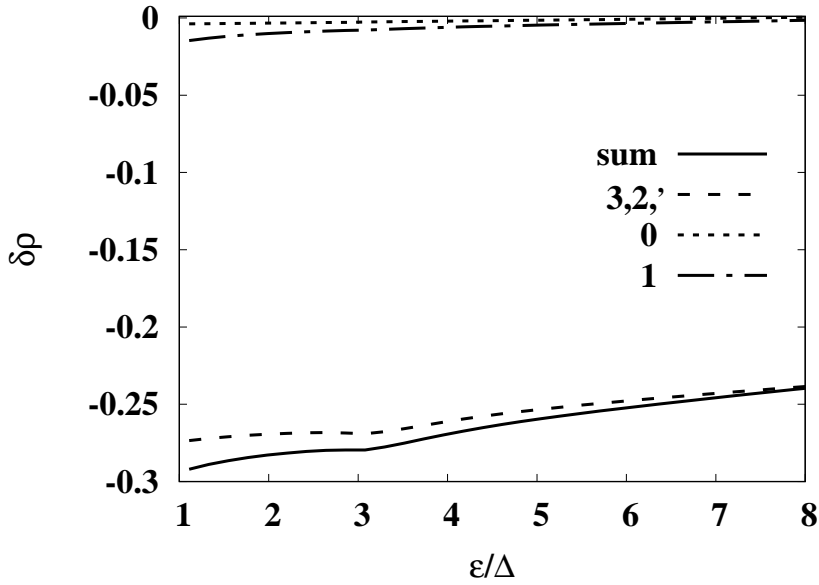}
\caption{\label{fig:5}
  The decomposition of $\delta\rho_{\epsilon}$
  into several terms according to $\Gamma_i$ and $\Gamma'$
  contained in $\delta\rho_{\epsilon}$.
  ``$3,2,{}^,$'', ''$0$'' and ``$1$''
  correspond to the suffixes of $\Gamma_i$ and $\Gamma'$.
  ``sum'' indicates the summation of these three quantities.
  $\alpha=120\Delta$ ($k_Fl=2.5$) and $T=0$.
}
\end{figure}

The difference in magnitude between
$\Gamma_i$ and $\Gamma'$ shown in Fig. 2 is directly reflected 
in $\delta\rho_{\epsilon}$.
$\delta\rho_{\epsilon}$ in the superconducting state is
decomposed into several terms, and
the results are shown in Fig.~\ref{fig:5}.
The decomposition is done according to $\Gamma_i$
and $\Gamma'$ contained in Eq. (\ref{eq:egddoscorr}).  
For example, ``$3,2,{}^,$'' in Fig. 5 represents
the contribution 
from $\Gamma_3$, $\Gamma_2$ and $\Gamma'$
to $\delta\rho_{\epsilon}$.
The calculated results show that
the phase and density fluctuations 
majorly contribute to $\delta\rho_{\epsilon}$
because they contain
the long-range part ($\propto 1/q^2$).
The contribution from the amplitude fluctuation
is small, as illustrated in Fig. 2. 

\begin{figure}
\includegraphics[width=11.5cm]{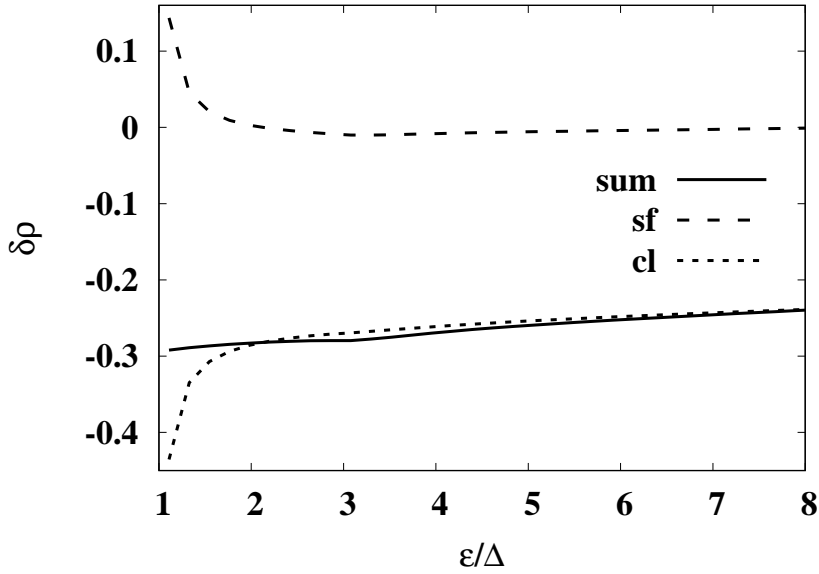}
\caption{\label{fig:6}
  The decomposition of $\delta\rho_{\epsilon}$.
  ``sf'' and ``cl'' indicate $\delta\rho^{sf}_{\epsilon}$
  and $\delta\rho^{cl}_{\epsilon}$, respectively.
    ``sum'' indicates the summation of these two quantities.
  $\alpha=120\Delta$ ($k_Fl=2.5$) and $T=0$.
}
\end{figure}

Equation (\ref{eq:egddoscorr}) 
seemingly includes a divergence 
proportional to $1/\sqrt{\epsilon^2-\Delta^2}$ in
$\delta\rho_{\epsilon}$.
To clarify the reason for the absence of this divergence
in Fig. 3,
we decompose Eq. (\ref{eq:egddoscorr})
as follows:
\begin{equation}
\rho'(\epsilon)
=\frac{\rho_0|\epsilon|}{\sqrt{\epsilon^2-\Delta^2}}
\left(\delta\rho^{sf}_{\epsilon}+\delta\rho^{cl}_{\epsilon}
\right)
\end{equation}
with
\begin{equation}
    \begin{split}
&\delta\rho^{sf}_{\epsilon}
=
   \frac{-3\sqrt{3\tau}}{(2\pi k_F l)^2}
   \int d\omega\int d x\sqrt{x} \\
&\times
   {\rm Im}
   \Bigl\{2{\rm coth}\left(\frac{\omega}{2T}\right)
\frac{{\rm Im}[\Gamma_i(q)](h_i+g^+_{\epsilon}g^+_{\epsilon-\omega}
       +h_i'f^+_{\epsilon}f^+_{\epsilon-\omega})+
2{\rm Im}[\Gamma'(q)](f^+_{\epsilon}g^+_{\epsilon-\omega}
            -g^+_{\epsilon}f^+_{\epsilon-\omega})}
          {(x+\zeta^+_{\epsilon}+\zeta^+_{\epsilon-\omega})^2}
                            \\
   &     +     {\rm tanh}\left(\frac{\epsilon-\omega}{2T}\right)
\frac{\Gamma_i(q)(h_i+g^+_{\epsilon}g^+_{\epsilon-\omega}
       +h_i'f^+_{\epsilon}f^+_{\epsilon-\omega})+
2\Gamma'(q)(f^+_{\epsilon}g^+_{\epsilon-\omega}
            -g^+_{\epsilon}f^+_{\epsilon-\omega})}
          {(x+\zeta^+_{\epsilon}+\zeta^+_{\epsilon-\omega})^2}\Bigr\}
    \end{split}
    \label{eq:egdsfdoscorr}
\end{equation}
and
\begin{equation}
    \begin{split}
&\delta\rho^{cl}_{\epsilon}
=
   \frac{-3\sqrt{3\tau}}{(2\pi k_F l)^2}
   \int d\omega\int d x\sqrt{x} \\
&\times
   {\rm Im}
   \Bigl\{
    {\rm tanh}\left(\frac{\epsilon-\omega}{2T}\right)
\frac{\Gamma_i(q)(h_i+g^+_{\epsilon}g^-_{\epsilon-\omega}
       +h_i'f^+_{\epsilon}f^-_{\epsilon-\omega})+
2\Gamma'(q)(f^+_{\epsilon}g^-_{\epsilon-\omega}
            -g^+_{\epsilon}f^-_{\epsilon-\omega})}
          {(x+\zeta^+_{\epsilon}+\zeta^-_{\epsilon-\omega})^2}\Bigr\}.
    \end{split}
    \label{eq:egdcldoscorr}
\end{equation}
The calculated results for these quantities
are shown in Fig.~\ref{fig:6}.
When $\Delta=0$,
Eqs. (\ref{eq:egdsfdoscorr}) and (\ref{eq:egdcldoscorr})
reduce to Eqs. (\ref{eq:nmsfdoscorr}) and (\ref{eq:nmcldoscorr})
(except for the factor $\rho_0$), respectively. 
Both $\delta\rho^{sf}_{\epsilon}$
and $\delta\rho^{cl}_{\epsilon}$ include
the effects of
the superconducting fluctuation and the Coulomb interaction
in the case of $\Delta\ne 0$.
$\delta\rho^{sf(cl)}_{\epsilon}$
includes only the ``retarded (advanced)''
quantities ($\zeta^{+(-)}_{\epsilon-\omega}$,
$g^{+(-)}_{\epsilon-\omega}$ and $f^{+(-)}_{\epsilon-\omega}$).
The calculated results show that
the absence of the divergence proportional to
$1/\zeta^+_{\epsilon}=i/\sqrt{\epsilon^2-\Delta^2}$
in $\delta\rho_{\epsilon}$
is caused by the cancellation between
the retarded and the advanced parts
[terms proportional to $1/
(x+\zeta^+_{\epsilon}+\zeta^+_{\epsilon-\omega})^2$
and $1/
(x+\zeta^+_{\epsilon}+\zeta^-_{\epsilon-\omega})^2$].

\begin{figure}
\includegraphics[width=11.5cm]{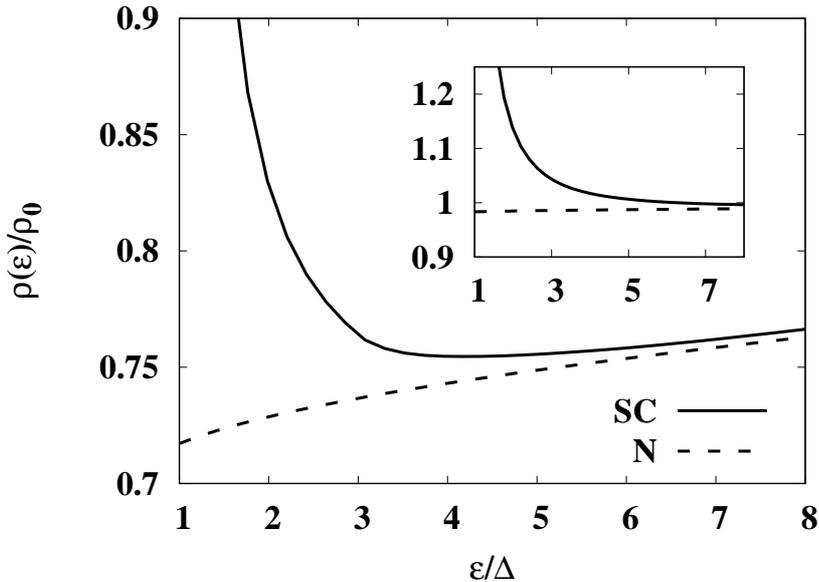}
\caption{\label{fig:7}
  The DOS with the correction included.
  $\alpha=120\Delta$ ($k_Fl=2.5$).
  ``SC'' and ``N'' indicate the result for the superconducting
  and the normal state, respectively.
  The inset shows the result for $\alpha=30\Delta$ ($k_Fl=10.0$).
}
\end{figure}

The DOS with the correction included
is written as follows:
\begin{equation}
  \rho(\epsilon)=\frac{\rho_0|\epsilon|}{\sqrt{\epsilon^2-\Delta^2}}
  +\rho'(\epsilon)=
  \frac{\rho_0|\epsilon|(1+\delta\rho_{\epsilon})}
       {\sqrt{\epsilon^2-\Delta^2}}.
\end{equation}
The calculated result of this expression 
is shown in Fig.~\ref{fig:7}.
In the normal state, $\rho(\epsilon)=\rho_0(1+\delta\rho_{\epsilon})$.
The result shows that $\rho(\epsilon)$ increases with
increasing $\epsilon$ for large $\alpha$.
For small $\alpha$, $\rho(\epsilon)$ decreases
as $|\epsilon|/\sqrt{\epsilon^2-\Delta^2}$ because of
the small values of $\delta\rho_{\epsilon}$.
This indicates that,
although the $\epsilon$-dependence of $\delta\rho_{\epsilon}$
is almost independent of $\alpha$, as shown in
Fig. 4,
the increasing DOS with $|\epsilon|$ is
observable only for large $\alpha$.

\section{Summary and Discussion}

In this study, we calculated the correction to
the DOS perturbatively.
The correction term is given by the Coulomb interaction
and the electron-phonon interaction, 
with vertices of these interactions modified by
the impurity scattering.
The modification enhances these interactions at low energy.
The energy dependence of the correction to DOS  
in the superconducting state 
is different from  that in the normal state,
and a dip structure appears at low energy. 
This structure is caused by the 
interaction which has a peak at about
twice the energy of the superconducting gap.
(The dip structure in the one-particle spectrum is
also observed in cuprates,
but its origin is different.~\cite{norman,dahm,takimoto})

There are two differences between the superconducting
state and the normal state.
First, the diffuson is modified because 
the opening of the superconducting gap changes
the dispersion of quasiparticles.
This gives rise to another pole in the diffuson propagator,
and this pole is treated correctly by
including the coupling of the density and phase fluctuations.
Second, the correction to DOS does not affect
the gap-edge singularity in the superconducting state.
This is because the 
cancellation between the retarded and advanced parts
occurs around the gap edge.
In the normal state, the superconducting fluctuation 
and the Coulomb interaction separately contribute
to the retarded and advanced parts, respectively.
In the superconducting state we cannot treat them
separately and need to include
both parts simultaneously in the correction to DOS.

Regarding the validity of perturbation expansion,
if we consider the perturbation expansion in the case of
the scattering by nonmagnetic impurities,
the correction to DOS is proportional to
${\rm Im}\sum_{{\mib k},k'}{\rm Tr}[\hat{G}_k\hat{\tau}_3
  \hat{G}_{k'}\hat{\tau}_3\hat{G}_k]=0$.
The nonmagnetic impurities do not affect
the DOS in the Born approximation.
In contrast, for paramagnetic impurities,
the correction to DOS is proportional to
${\rm Im}\sum_{{\mib k},k'}{\rm Tr}[\hat{G}_k\hat{\tau}_0
  \hat{G}_{k'}\hat{\tau}_0\hat{G}_k]\propto
\Delta^2|\epsilon|/(\epsilon^2-\Delta^2)^{3/2}$.
This means that the perturbation expansion is invalid
around $|\epsilon|\simeq \Delta$,
and the gap edge in the DOS
changes qualitatively.~\cite{abrikosov61,skalski}
The calculation in this paper shows that
the correction to DOS does not diverge around the gap edge.
This indicates that the perturbation expansion is
valid within our approximations.

We calculated the Fock term
with its vertices modified by diffuson
(for example, Fig. 3 (a) in Ref. 6,
with
the wavy line in this figure replaced by
the Coulomb interaction and the superconducting fluctuation
in our calculation).
It is possible to consider other types of diagrams. 
For example, these are the Fock terms with its vertices 
modified by Cooperon and the Hartree term
(Figs. 3 (b)$-$(d) in Ref. 6
).
The correction to DOS
by the Fock term with Cooperon
is proportional to
$-{\rm Im}\sum_{{\mib k},q}{\rm Tr}[
  \hat{G}_k\cdots
  \sum_{k_1,k_2}
  \Gamma_{k_1-k_2}\hat{G}_{k_1}\hat{\tau_3}\hat{G}_{k_2}
  \cdots\hat{G}_{q-k}\cdots\hat{G}_{q-k_1}
  \hat{\tau_3}\hat{G}_{q-k_2}\cdots
  \hat{G}_k]$.
The singular part $\Gamma_{q}\propto 1/{\mib q}^2$ (which
majorly contributes to the correction to DOS in our calculation)
is weakened
when the summations are performed.
Thus, we can omit this type of diagram.
There is a similar term
in the case of the Hartree diagram
modified by diffuson or Cooperon.
(In the case of the Fock term modified by diffuson,
$-{\rm Im}\sum_{{\mib k},q}{\rm Tr}[
  \hat{G}_k\cdots
  \sum_{k_1,k_2}\Gamma_{q}\hat{G}_{k_1}\hat{\tau_3}\hat{G}_{k_1-q}
  \cdots\hat{G}_{k-q}\cdots\hat{G}_{k_2-q}
  \hat{\tau_3}\hat{G}_{k_2}\cdots
  \hat{G}_k]$.)

This study considers the case of low temperatures
($T\ll\Delta$).
The superconducting gap
$\Delta$ was taken as the unit of energy, and
we did not consider the interaction effect on the superconducting gap.
When the temperature is comparable to the superconducting gap,
the self-consistency through the gap equation
becomes important.

Finally, we comment on the possibility of
observing a dip structure in experiments.
Experimentally, it is known that
the superconducting state becomes
inhomogeneous with decreasing $k_F l$,~\cite{chand}
and the one-particle spectrum is averaged
over these inhomogeneous states.
(There are also theoretical studies
on inhomegeneities in superconductors
without Coulomb interaction.~\cite{larkin,ghosal}
In addition, the perturbative calculation should be
modified for 
small values of $k_F l$ near the insulating state,
and the renormalization-group method~\cite{burmistrov}
will be required.)
Thus, it is difficult to observe the dip structure
in the case of large values of $\alpha$.
Figures 4 and 7 show, however, that
the dip structure is possibly observed
even for small values of $\alpha$ ($k_Fl\gg 1$,
but in the dirty limit $\Delta\tau\ll 1$)
when the overall factor $|\epsilon|/\sqrt{\epsilon^2-\Delta^2}$
is removed.
  The dip structure originates from the interactions
  in the superconducting state, and therefore 
  the difference between our calculation and the calculations
  using the Coulomb interaction and diffuson of the normal
  state~\cite{browne,rabatin} appears in this quantity.

\section*{Acknowledgment}

The numerical computation in this work was carried out 
at the Yukawa Institute Computer Facility.


\begin{thebibliography}{9}





\bibitem{anderson59} P. W. Anderson, 
J. Phys. Chem. Solids {\bf 11}, 26 (1959). 

\bibitem{schmid} A. Schmid,
  Z. Physik {\bf 271}, 251 (1974).

\bibitem{altshulerSSC1979} B. L. Altshuler and A. G. Aronov,
  Solid State Commun. {\bf 30}, 115 (1979).

\bibitem{altshulerJETP1979} B. L. Al'tshuler and A. G. Aronov,
  Sov. Phys. JETP {\bf 50}, 968 (1979).

\bibitem{ovchinnikov}   Yu. N. Ovchinnikov,
  Sov. Phys. JETP {\bf 37}, 366 (1973).
  
\bibitem{maekawa} S. Maekawa and H. Fukuyama,   J. Phys. Soc. Jpn. {\bf 51}, 1380 (1981).

\bibitem{takagi} H. Takagi and Y. Kuroda,
  Solid State Commun. {\bf 41}, 643 (1982).



\bibitem{altshulerPRL1980} B. L. Altshuler, A. G. Aronov, and P. A. Lee,
  Phys. Rev. Lett. {\bf 44}, 1288 (1980).


\bibitem{fukuyama} H. Fukuyama,
  J. Phys. Soc. Jpn. {\bf 48}, 2169 (1980).

  


\bibitem{abrahams} E. Abrahams, P. W. Anderson, P. A. Lee, and
  T. V. Ramakrishnan, 
  Phys. Rev. B {\bf 24}, 6783 (1981).





\bibitem{abrahams1970} E. Abrahams, M. Redi, and J. W. F. Woo,
    Phys. Rev. B {\bf 1}, 208 (1970).


\bibitem{dicastro} C. Di Castro, R. Raimondi, C. Castellani, and A. A. Varlamov,
Phys. Rev. B {\bf 42}, 10211 (1990).



\bibitem{sacepe2008} B. Sac\'ep\'e, C. Chapelier, T. I. Baturina, V. M. Vinokur, M. R Bakanov, and M. Sanquer,
  Phys. Rev. Lett. {\bf 101}, 157006 (2008).  


\bibitem{carbillet2016} C. Carbillet, S. Caprara, M. Grilli, C. Brun, T. Cren, F. Debontridder, B. Vignolle, W. Tabis, D. Demaille, L. Largeau, K. Ilin, M. Siegel, D. Roditchev, and B. Leridon,
  Phys. Rev. B {\bf 93} 144509 (2016).


\bibitem{kamlapure} A. Kamlapure, T. Das, S. C. Ganguli, J. B Parmar, S. Bhattacharyya, and P. Raychaudhuri,
Sci. Rep. {\bf 3}, 2979 (2013)





\bibitem{chockalingam} S. P. Chockalingam, M. Chand, A. Kamlapure, J. Jesudasan, A. Mishra, V. Tripathi, and P. Raychaudhuri,
  Phys. Rev. B {\bf 79} 094509 (2009).


\bibitem{chand} M. Chand, G. Saraswat, A. Kamlapure, M. Mondal, S. Kumar, J. Jesudasan, V. Bagwe, L. Benfatto, V. Tripathi, and P. Raychaudhuri,
Phys. Rev. B {\bf 85}, 014508 (2012).



\bibitem{browne} D. A. Browne, K. Levin, and K. A. Muttalib,
  Phys. Rev. Lett. {\bf 58}, 156 (1987).

    
\bibitem{rabatin} B. Rabatin and R. Hlubina, 
  Phys. Rev. B {\bf 98}, 184519 (2018).






\bibitem{anderson58}   P. W. Anderson,
  Phys. Rev. {\bf 112}, 1900 (1958). 

\bibitem{abrikosov59} A. A. Abrikosov and L. P. Gor'kov, 
Sov. Phys. JETP {\bf 8}, 1090 (1959). 

\bibitem{AGD} A. A. Abrikosov, L. P. Gor'kov, and I. E. Dzyaloshinskii, 
  {\it Methods of Quantum Field Theory in Statistical Physics} (Pergamon, Oxford, 1965) Chap. 7, Sec. 37.1.

\bibitem{norman} M. R. Norman, H. Ding, J. C. Campuzano, T. Takeuchi, M. Randeria, T. Yokoya, T. Takahashi, T. Mochiku, and K. Kadowaki,
  Phys. Rev. Lett. {\bf 79} 3506 (1997).


\bibitem{dahm} T. Dahm, D. Manske, and L. Tewordt,
  Phys. Rev. B {\bf 58} 12454 (1998).

  
\bibitem {takimoto} T. Takimoto and T. Moriya,
  J. Phys. Soc. Jpn. {\bf 67} 3570 (1998).


\bibitem{abrikosov61} A. A. Abrikosov and L. P. Gor'kov, 
Sov. Phys. JETP {\bf 12}, 1243 (1961). 

\bibitem{skalski} S. Skalski, O. Betbeder-Matibet, and P. R. Weiss, 
  Phys. Rev. {\bf 136}, A1500 (1964).
  
\bibitem{larkin} A. I. Larkin and Yu. N. Ovchinnikov,
  Sov. Phys. JETP {\bf 34}, 1144 (1972).  

  
\bibitem{ghosal} A. Ghosal, M. Randeria, and N Trivedi,
Phys. Rev. B {\bf 65}, 014501 (2001).


\bibitem{burmistrov} I. S. Burmistrov, I. V. Gornyi, and
  A. D. Mirlin,
Phys. Rev. B {\bf 93}, 205432 (2016).


  





\end{thebibliography}
\end{document}